# Origin of Ferromagnetism in nitrogen embedded ZnO:N thin films

Chang-Feng Yu, Tzu-Jen Lin
Department of Applied Physics, National Chiayi University, Chiayi 60004, Taiwan
E-mail: cfyu@mail.ncyu.edu.tw
Shih-Jye Sun
Department of Applied Physics, National University of Kaohsiung, Kaohsiung 811, Taiwan
Hsiung Chou
Department of Physics and Center for Nanoscience and Nanotechnology, National Sun Yat-Sen University, Kaohsiung 804, Taiwan

**Abstract**: Nitrogen embedded ZnO:N films prepared by pulsed laser deposition exhibit significant ferromagnetism. The nitrogen ions contained in ZnO confirmed by Secondary Ion Microscopic Spectrum and Raman experiments and the embedded nitrogen ions can be regarded as defects. According to the experiment results, a mechanism is proposed based on one of the electrons in the completely filled d-orbits of Zn that compensates the dangling bonds of nitrogen ions and leads to a net spin of one half in the Zn orbits. These one half spins strongly correlate with localized electrons that are captured by defects to form ferromagnetism. Eventually, the magnetism of nitrogen embedded ZnO:N films could be described by a bound magnetic polaron model.

## 1. Introduction

Magnetic ion doped transition metal oxides such as ZnO, $TiO_2$, $SnO_2$, $In_2O_3$ and $HfO_2$ are candidates for room temperature diluted magnetic semiconductors [1-4] with highly industrial applications in spintronics. The most popular material used is ZnO, which has been investigated in recent years as a transparent conducting oxide (TCO) because of its good electrical and optical properties in combination with a large band gap of 3.3 eV, abundance in nature, and absence of toxicity with a wide variety of applications as electrodes, window materials in displays, solar cells, and various optoelectronic devices [5-8]. Various techniques have been used to deposit doped ZnO films on different substrates, including Metal-Organic Chemical Vapor Deposition (MOCVD), Molecular Beam Epitaxy (MBE), Pulsed Laser Deposition (PLD), and Spray Pyrolysis Deposition (SPD). Furthermore, compared with other deposition methods, PLD is characterized by several advantages, such as low substrate temperatures, good adhesion on substrates, and the fact that alloys and compounds of materials with different vapor pressures can be deposited easily. The possible mechanism for magnetic ion doped ferromagnetic oxides has been proposed [9], called the bound magnetic polaron (BMP) model, which is different from the conducting carriers mediated mechanism of RKKY [10] appearing in III-V diluted magnetic semiconductors [11]. The BMP model is based on the donor electrons being polarized by the magnetic spins of doped ions to form bond states with finite orbital radius.

When the dopant density attains a critical value (percolation), the neighboring orbits overlap to create a spin-split impurity band leading to interesting conducting electricity and magnetism. Recent experiments demonstrate that some undoped and nitrogen embedded transition metal oxides also exhibit ferromagnetism [12,13]. Similarly, our N embeded ZnO:N films prepared by PLD in a $N_2O$ partial pressure atmosphere show significant ferromagnetism, similar to the $TiO_2$:N films. A perfect ZnO and $TiO_2$ samples should not show magnetism because the d-orbits of Zn are completely filled. The picture of the mechanism for the undoped transition metal oxides could be different from BMP because there are no magnetic ions doped directly. Therefore, it is an interesting and physically rich problem to study the mechanism of magnetism for the nitrogen (N) embedded ZnO films with no doping of access magnetic ions.

## 2. Experiment detail and analyses

A reliable method was used to deposit the thin films of undoped zinic oxide transparent electrodes on glass substrates by PLD. The ablation PLD targets with dimensions of 1 inch diameter x 0.125 inch thick were fabricated with high purity (5N) powder by the solid state reaction technique. ZnO:N films were deposited at 150$^o$C in Oxygen or $N_2O$ (99.99%) atmosphere of 150 mTorr on the glass substrates. A double frequency with Q switched Nd:YAG pulsed laser operated at 532 nm, pulse duration of about 7 ns, and a 18.5 mJ/cm2 energy density was focused on the target to generate plasma plume. The target to substrate distance was 5.0 cm. The crystallinity and surface morphology of ZnO films were characterized by x-ray diffraction (Rigaku Multiflex CD3684N diffractometer). The XRD data for undoped and ZnO:N films grown at the same substrate temperatures are represented in Fig. 1. Both films exhibit a highly preferred (002), 34.2$^o$, orientation indicating that ZnO prepared by PLD show a good textured growth with c-axis perpendicule to the substrate that is similar to others reports [14-19]. Inspection of Raman shift is an indirect method to probe the possible embedded nitrogen in our films. It has been proposed by Wang *et. al.* that the access peaks at 275 and 582 cm$^{-1}$ in an unpolarized Raman spectrum are highly possible caused by the existence of embedded N in ZnO structure [20].

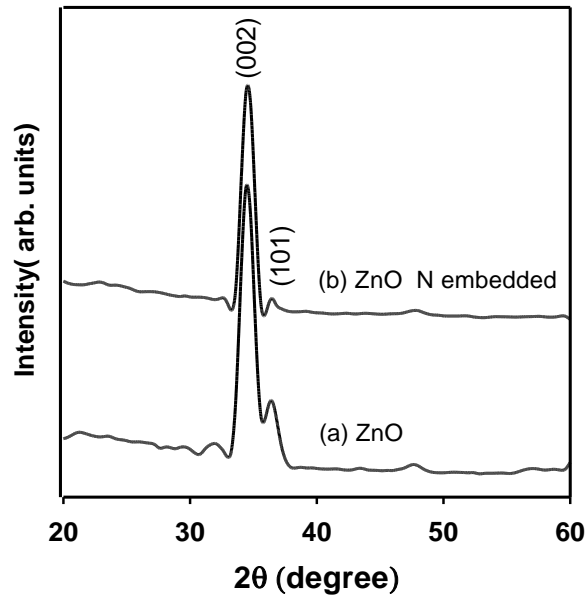

Fig. 1 X-ray diffraction spectra of (a) undoped ZnO and (b) N-embedded ZnO:N thin films onto glass substrate showing highly preferred (002) orientation.

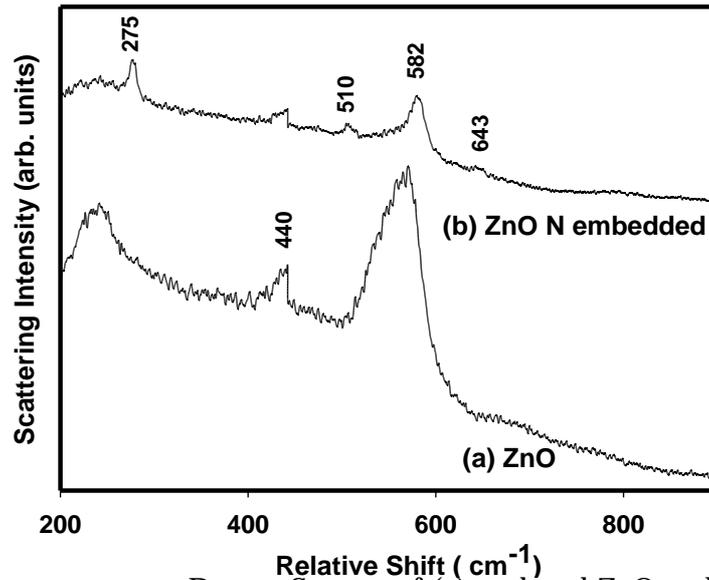
Fig. 2 Room temperature Raman Spectra of (a) undoped ZnO and (b) N-embedded ZnO:N films onto glass substrate. The additional modes in N-embedded ZnO thin films are marked by their respective frequencies.

In Fig. 2, two peaks at 440 and 580 cm$^{-1}$ are those of ideal ZnO structure while the other two peaks at 275 and 582 cm$^{-1}$ coincide with Wang's speculation. Secondary Ion Microscopic Spectrum (SIMS) was taken for a 325Å N-embedded ZnO film, as shown in Fig. 3, in which the concentration of oxygen is low that indicating a high level of oxygen vacancies was created during the film growth, and that a low concentration of embedded nitrogen was detected. The compositional profile of Zn, O and N are distributed in uniform throughout the entire film.

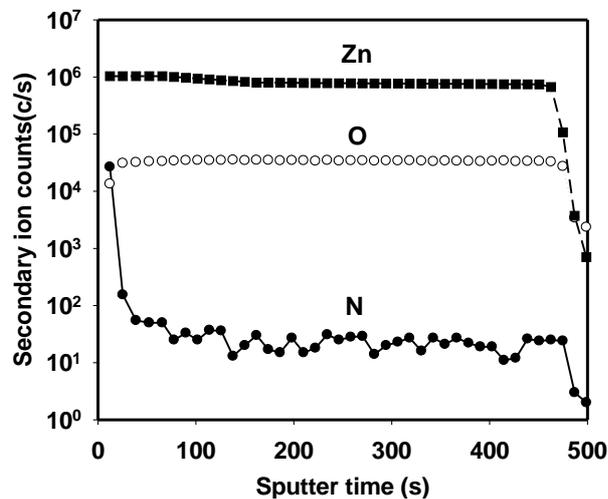
Fig. 3. SIMS depth profile for ZnO:N thin film

The photoluminescence was measured with a He-Cd laser as a light source using an excitation wavelength of 325 nm. Figure 4 is the comparison of PL spectra of the undoped and N-embedded ZnO films. All of the spectra consist of two majority bands. The first one, centered at photon energy of 3.28 eV for ultraviolet band, corresponds to the radiative recombination of excitons. The second PL bands show a broad orange emission band with maximum in the range between 2.0 and 2.4 eV. It is possible that the introduction of the impurity atoms changes the potential energy around the impurity, and results in the variation of the energy levels of oxygen vacancies. The radiative recombination process responsible for the orange emission would take place between the donors associated with oxygen vacancies and the acceptors associated with the native defects adjacent to the impurity [24-27]. When $N_2O$ dopant is introduced, the PL spectra of N-embedded ZnO thin film show that the formation of ultraviolet band is suppressed as indicated in Figure 4 (b). The suppression of ultraviolet band could be attributed to the combination with the oxygen species in the films and/or the occupation of N dopant in the interstitial sites. This implies that the N dopants in our films actually behave as donors. The magnetization versus magnetic field at room temperature for a 60 nm-thick pure ZnO film is shown in Figure 5. One can see that the film is ferromagnetic at room temperature. The observed magnetism in ZnO films is unexpected, because neither $Zn^{2+}$ nor $O^{-2}$ is magnetic; thus, in principle, there is no source for magnetism in ZnO. From the electrical results of CAFM (Conducting AFM) experiments, the conductivity of the nitrogen embedded ZnO films decreases compared with pure ZnO films as shown in Fig. 6. This means that the embedded N ions diminish the transport carrier density. It is reasonable to propose that these embedded nitrogen ions create defects, which capture part of itinerant electrons. It is based on this capture evidence that we propose the magnetic mechanism described in next section.

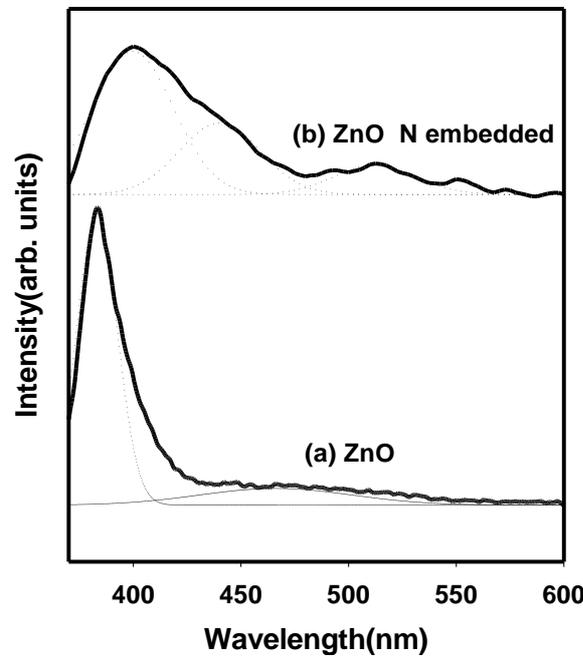

Fig. 4 Normalized photoluminescence spectra for (a) undoped ZnO and (b) N-embedded ZnO:N thin films at room temperature.

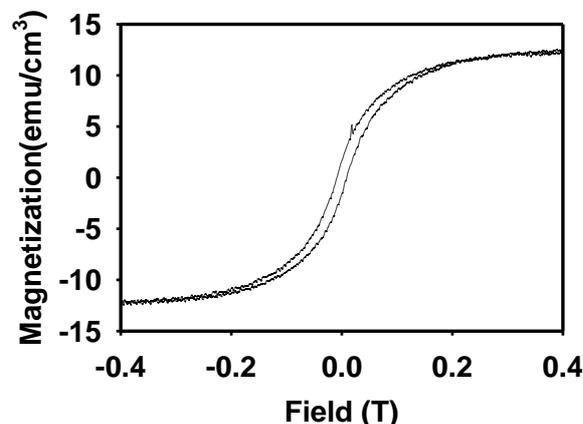

Fig. 5 Magnetization versus magnetic field at 300 K for the N-embedded

ZnO:N thin film. H was applied parallel to the film plane. Signals from the substrates were subtracted.

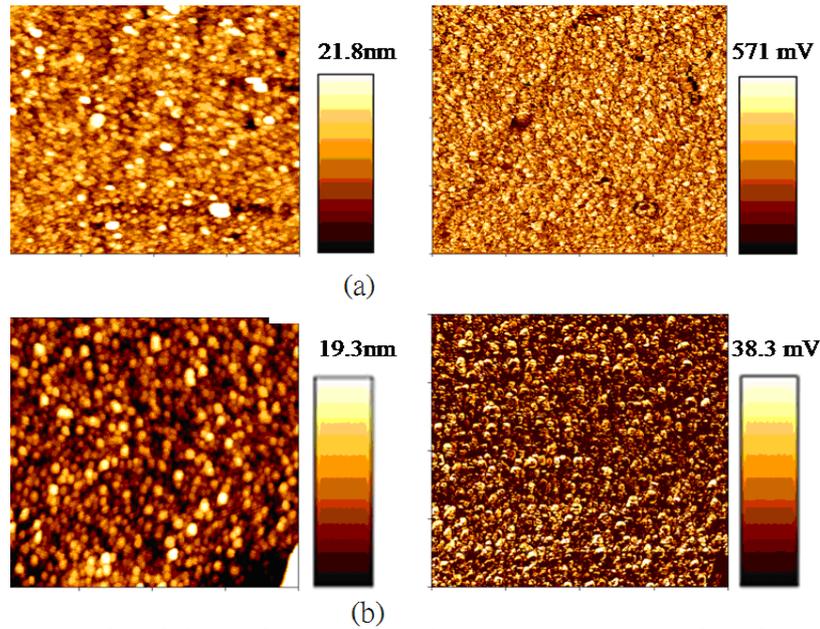

(a)

(b)

Fig. 6 Topography (left) and current (right) images of (a) undoped ZnO and (b) N-embedded ZnO:N films onto glass substrate. The tip was biased to -0.5V and the ZnO thin film was grounded.

## 3. Model construction

In the BMP model for transition metal oxides, the defect comes from the oxygen vacancy which could catch one itinerant electron to form a hydrogen-like orbit with a finite radius. When the defects increase to a critical amount, the orbits overlap to form a narrow impurity band. The doped magnetic ions within the radius coverage and correlate through the impurity band electrons to become ferromagnetic. Obviously, the important element of BMP, the magnetic ion, does not appear in our ferromagnetic samples. However, we believe the magnetism of undoped ZnO eventually should come from the net spins in d-orbits. Based on the SIMS and Raman experiment, the N is embedded in our samples. It is seen that, for the valence charges of N, O, Zn and ZnO, every N ion beside a Zn ion leaves an uncompensated dangling bond. In order to stabilize the system, these dangling bonds should be compensated by capturing electrons. When the ion N is beside Zn, one electron in the completely filled d-orbits of Zn possibly jumps to compensate the dangling bond to reduce the total energy leading to a net spin with one half in the d-orbit of Zn and leads to a possibility of ferromagnetism; this picture is depicted in Fig. 6. Eventually, these nitrogen embedded ZnO:N films, after the intra carriers jump between d-orbits of Zn and defects, are equivalent to the doped ZnO samples.

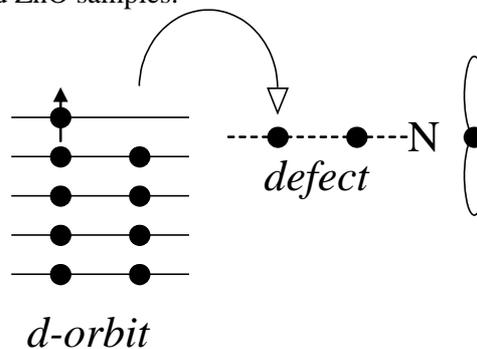

*d-orbit*

Fig. 6 The picture of the magnetic model for N-embedded ZnO: N one of the electrons in the completely filled d-orbits of Zn leaves to compensate

the dangling bonds of N ions beside Zn ions leading to a net spin one half in Zn orbits.

## 4. Conclusion

In Conclusion, our nitrogen embedded undoped ZnO shows significant ferromagnetism. The conductivity measured by AFM represents that the itinerant electrons are captured by defects induced by nitrogen ions. According to the experiment results, a model of defect induced magnetism is proposed based on one of the electrons in the completely filled d-orbits of Zn that compensates the dangling bonds of nitrogen ions and leads to a net spin of one half in the Zn orbits.

**Acknowledgments**
This work was supported by the National Science Council in Taiwan through Grant No. NSC-95-2112-M-390-002-MY3 (SJ. Sun), and No. NSC-95-2112-M-110-011-MY3 (H. Chou). We appreciate professors C. D. Hu and D. J. Huang and Lance Horng for their useful discussions.

**Reference**
[1]. Venkatesan M, Fitzgerald C B, Lunney J G and Coey J M D, Phys. Rev. Lett. 93, 177206 (2004)
[2]. Hong N H, Sakai J, Prellier W, Hassini A, Ruyter A and Gervais F, Phys. Rev. B 70, 195204 (2004)
[3]. Ogale S B, Choudhary R J, Buban J P, Lofland S E, Shinde S R, Kale S N, Kulkarni V N, Higgins J, Lanci C, Simpson J R, Browning N D, Das Sarma S, Drew H D, Greene R L and Venkatesan T, Phys. Rev. Lett. 91, 77205 (2003)
[4]. He J, Xu S, Yoo Y K, Xue Q, Lee H C, Cheng S, Xiang X-D, Dionne G and Takeuchi I, Appl. Phys. Lett. 86, 052503 (2005)
[5]. K. Vanheusden, W.L. Warren, C. H. Seager, D. R. Tallant, J.A. Voight and B.E. Gnade, J. Appl. Phys 79, 7983 (1996)
[6]. C.J. Sheppard, Thesis: Structural and Optical Characterization of α Si:H and ZnO; Faculty of science at the Rand Afrikaans University, (2002)
[7]. D.P. Norton, Y. W. Heo, M.P. Ivill, K. Ip, S. J. Pearton, M. F. Chisholm and T. Steiner, Materials today, 7, 34 (2004)
[8]. H.Y. Kim, J.H. Kim, Y.J. Kim, K.H. Chae, C.N. Whang, J.H. Song and S. Im, Optical materials 17, 141 (2001)
[9]. J. M. D. Coey, M. Venkatesan and C. B. Fitzgerald, Nature materials, 4, 173 (2005)
[10]. C. Kittel, see the Chaper 18 in "Quantum Theory of Solids".
[11]. H. Ohno, Science 281, 951 (1998)
[12]. Nguyen Hoa Hong, Joe Sakai and Virginie Brizé, J. Phys.:Cond. Matt. 19, 036219 (2007)
[13]. Soack Dae Yoon, Yajie Chen, Aria Yang, Trevor L Goodrich, Xu Zuo, Dario A Arena, Katherine Ziemer, Carmine Vittoria and Vincent G Harris, J. Phys.:Cond. Matt. 18, L355 (2006)
[14]. S.H Jeong, S. Kho, D. Jung, S.B. Lee, J.H. Boo, Surface and coatings technology, 174-175, 187-192 (2003)
[15]. S. Bose, S. Ray and K. Barua, J. Phys D: Appl. Phys, 29, 1873 (1996).
[16]. K. Ho Kim, K. Cheol Park and D. Young Ma, J. Appl. Phys. 81, 7764 (1997)
[17]. R. Cebulla, R. Wendt and K. Ellmer, J. Appl. Phys. 83, 1087 (1998)
[18]. K. Ellmer, J. Phys D: Appl. Phys, 33, R17-R32 (2000)
[19]. D. J. Kang, J.S. Kim, S.W. Jeong, Y. Roh, S.H. Jeong, J.H. Boo, Thin solid films 475, 160 (2005).
[20]. T. C. Damen, S. P. S. Porto, and B. Tell, Phys. Rev. 142, 570 (1966)
[21]. X. Wang, S. Yang, J. Wang, M. Li, X. Jiang, G. Du, X. Liu, and R. P. H. Chang, J. Crys. Grow. 226, 123 (2001)

[22]. Lei L. Kerr, Xiaonan Li, Marina Canepa, Andre J. Sommer, Thin Solid Films. 515, 5282 (2007)
[23]. A. Kaschner, U. Haboeck, Martin Strassburg, Matthias Strassburg, G. Kaczmarczyk, A. Hoffmann, and C. Thomsen, Appl. Phys. Lett. 80, 1909 (2002)
[24]. S. B. Zhang, S.-H. Wei, and Alex Zunger, Phys. Rev. B 63, 075205 (2001)
[25]. M. Liu, A. H. Kitai, P. Mascher, Journal of Luminescence, 54, 35 (1992)
[26]. K. Vanheusden, C. H. Seager, W. L. Warren, D. R. Tallant, J. A. Voigt, Appl. Phys. Lett. 68, 403 (1996).
[27]. Bixia Lin, Zhuxi Fu, and Yunbo Jia, Appl. Phys. Lett. 79, 943 (2001)